\documentclass[groupedaddress, superscriptaddress, twocolumn, pra,nofootinbib]{revtex4-2}
\usepackage{graphicx}
\usepackage{calc}
\usepackage{amsmath}
\usepackage{amsthm}
\usepackage{amssymb}
\usepackage{bm}
\usepackage{color}
\usepackage{lipsum}
\usepackage{multirow}
\usepackage[pdftex,dvipsnames,usenames]{xcolor}
\usepackage[colorlinks=true,urlcolor=blue,citecolor=blue,linkcolor=blue]{hyperref}

\DeclareMathOperator{\Tr}{Tr}

\newcommand{\ket}[1]{\left|#1\right\rangle}
\newcommand{\bra}[1]{\left\langle#1\right|}
\DeclareMathOperator{\Haf}{Haf}
\DeclareMathOperator{\Tor}{Tor}
\DeclareMathOperator{\Ken}{Ken}
\DeclareMathOperator{\diag}{diag}
\usepackage{amsfonts}

\begin{document}
    \title{Realistic photon-number resolution in Gaussian boson sampling}
	
	\author{I. S. Yeremenko}
	\affiliation{Bogolyubov Institute for Theoretical Physics, NAS of Ukraine, Vulytsia Metrologichna 14b, 03143 Kyiv, Ukraine}
	
	\author{M. A. Dmytruk}
	\affiliation{Physics Department, Taras Shevchenko National University of Kyiv, Prospect Glushkova 2, 03022 Kyiv, Ukraine}
    \affiliation{Bogolyubov Institute for Theoretical Physics, NAS of Ukraine, Vul. Metrologichna 14b, 03143 Kyiv, Ukraine}
	
	\author{A. A. Semenov}
	\affiliation{Bogolyubov Institute for Theoretical Physics, NAS of Ukraine, Vul. Metrologichna 14b, 03143 Kyiv, Ukraine}
	\affiliation{Department of Theoretical and Mathematical Physics, Kyiv Academic University, Boulevard Vernadskogo  36, 03142  Kyiv, Ukraine}
	
\begin{abstract}
Gaussian boson sampling (GBS) is a model of nonuniversal quantum computation that claims to demonstrate quantum supremacy with current technologies.
This model entails sampling photocounting events from a multimode Gaussian state at the outputs of a linear interferometer. 
In this scheme, collision events---those with more than one photon for each mode---are infrequent. 
However, they are still used for validation purposes.
Therefore, the limitation of realistic detectors to perfectly resolve adjacent photon numbers becomes pivotal.
We derive a the photocounting probability distribution in GBS schemes which is applicable for use with general detectors and photocounting techniques.
This probability distribution is expressed in terms of functionals of the field-quadrature covariance matrix, e.g., Hafnian and Torontonian in the well-known special cases of photon-number resolving and on-off detectors, respectively. 
Based on our results, we consider a GBS validation technique involving detectors with realistic photon-number resolution.
\end{abstract}
\maketitle

%%%%%%%%%%%%%%%%%%%%%%%%%%%%%%%%%%%%%%%%%%%%%%%%%%%%%%%%%%%%%%%%%%%%%%%%%%%
%%% Introduction
%%%%%%%%%%%%%%%%%%%%%%%%%%%%%%%%%%%%%%%%%%%%%%%%%%%%%%%%%%%%%%%%%%%%%%%%%%%

\section{Introduction}

Since the time when Aaronson and Arkhipov proposed the boson sampling (BS) model and demonstrated its computational complexity \cite{aaronson2013}, significant progress has been made towards its experimental implementation. 
A primary reason for this is the reformulation of the basic idea to Gaussian boson sampling (GBS) proposed in Ref.~\cite{Hamilton2017}.
A key distinction of GBS is that it uses nonclassical Gaussian states instead of single-photon states at the input of the linear inteferometer.
This replacement simplifies the experimental setup, making it feasible to create large-scale quantum devices that demonstrate quantum supremacy \cite{Zhong2020, Zhong2021,Madsen2022,deng2023}.
As experimental techniques have evolved, the issue of certification has increased in prominence.
However, in the case of BS and GBS, the classical data required for a direct comparison with the data generated by the quantum device are not available.
This makes full certification impossible.

Consequently, the emphasis has shifted from full certification to validation, aiming to exclude 
the possibility of replicating the generated data using various classical models. 
Most validation methods were initially developed for BS but have since been extended to encompass GBS. 
They include algorithms for approximate classical simulations of GBS \cite{clifford2017,Quesada2020, Quesada2022, Oh2022, Bulmer2022, Popova2022, cilluffo2023, Oh2024}.
Potential complications in experiments, such as photon distinguishability \cite{deng2023,Renema2018, Moylett2019, renema2020, Shi2021} and photon losses \cite{Oszmaniec2018, GarciaPatron2019, Brod2020,Qi2020, Oh2021, liu2023}, have been explored. 
These factors can affect the problem of complexity and might even enable classical simulation of the experiment \cite{Rahimi-Keshari2016,Oh2024}.

Beyond the question of what to validate lies the issue of how to perform this validation. 
Until recently, GBS experiments were conducted using a model proposed in Ref.~\cite{Quesada2018}, which lacks photon-number resolution at the output. 
This approach simplified experiments but, at the same time, limited the validation methods available to enhance our confidence in the experimental setup. 
Under these conditions, a number of validation methods have been proposed, for instance, tests based on Bayesian methods \cite{wu2021,spagnolo2014,MartinezCifuentes2023}, statistical properties of two-point correlation functions \cite{Walschaers2016, Phillips2019, Giordani2018} or higher-order correlations \cite{wu2021}, and grouped  \cite{Dellios2022, Drummond2022,dellios2023a,dellios2023b,bressanini2024} and marginal  \cite{renema2020, villalonga2022} probabilities.

The use of photon-number resolving (PNR) detectors in GBS experiments paves the way for an engaging class of validation methods.
For example, in Ref.~\cite{Giordani2023}, the authors raised questions about the validation of GBS with ideal PNR detectors based on components of graph feature vectors, known as orbits.
A GBS experiment with transition-edge sensors \cite{lita08,Arrazola2021}, which under ideal conditions discriminate between the numbers of photons up to a predetermined threshold, was reported in Ref.~\cite{Madsen2022}.   
In Ref.~\cite{deng2023} a GBS experiment with so-called click detectors (also called pseudo-PNR detectors) was presented (cf. Ref.~\cite{paul1996,castelletto2007,schettini2007,blanchet08,achilles03,fitch03,rehacek03}).
Although the photon-number resolution for this detection technique is imperfect (see Ref.~\cite{sperling12a}), such experiments still provide an opportunity to explore validation methods not accessible in the case of on-off detectors. 

Building ideal PNR detectors is a challenging task for currently available technologies.
In most realistic scenarios, the measurement outcomes of the detectors (the number of clicks, pulses, etc.) differ randomly from the number of received photons.
This means that the corresponding elements of the positive operator-valued measure (POVM) are not projectors on Fock states.
Therefore, it is important to consider a GBS model incorporating detectors with realistic photon-number resolution, which is considered for the BS model \cite{Len2022}.
In addition, the appropriate validation techniques should be reformulated for such detectors. 

Several widely-used experimental techniques enable an approximate resolution between adjacent numbers of photons.
The first one is related to the click detectors mentioned above.
In this case, the light beam is demultiplexed in several spatial \cite{paul1996,castelletto2007,schettini2007,blanchet08} or temporal \cite{achilles03,fitch03,rehacek03} modes, and each of them is analyzed with an on-off detector.
The outcome of such detectors corresponds to the number of triggered detectors (clicks).
The theoretical description of such detectors was developed in Ref.~\cite{sperling12a}.

Another technique is based on counting photocurrent pulses within a measurement time window.
In this case, the dead time of the detectors may significantly reduce the ability to resolve between numbers of photons, as is the case the avalanche photodiodes (APDs).
A theoretical description of this photodetection technique was presented in Refs.~\cite{ricciardi66,muller73,muller74,cantor75,teich78,vannucci78,rapp2019} and Ref.~\cite{semenov2023} for classical and quantum light, respectively.
When using superconducting nanowire single-photon detectors (SNSPDs) \cite{natarajan13, You2020, SEMENOV2001,Gol`tsman2001, Zhang2019, Zadeh2021}, one should additionally account for the relaxation time; see Ref.~\cite{Uzunova2022} for a theoretical description of photocounting measurements in this case.

In this paper we systematically consider the GBS model, accounting for photon-number resolution of realistic detectors.
First, we show that the photocounting distribution is expressed in terms of a matrix functional specific to each type of detection.
In particular, this functional reduces to well-known forms, e.g., Torontonian or Hafnian, for on-off and PNR detectors, respectively. 
Second, we tailor validation methods for GBS to the case of realistic photon-number resolution.

Many proposals for applications of GBS use the assumption of the ideal photon-number resolution.
It is  related, for example, to applications in graph theory \cite{Arrazola2018a,Bradler2018,bradler2019,Schuld2020, Bradler2021}, point processes \cite{Jahangiri2020}, quantum chemistry \cite{Huh2015}, molecular docking used for drug design \cite{Banchi2020}, etc.
However, imperfect photon-number resolution may significantly modify the outputs of these techniques.
Therefore, any practical application of GBS should take this aspect into account.

The rest of the paper is organized as follows.
In Sec.~\ref{Sec:Ph-count_Pr}, we introduce a universal formula for the photocounting distribution in the GBS model with realistic photon-number resolution.
A validation method tailored to GBS with realistic photon-number resolution is considered in Sec.~\ref{Sec:Orbit_pr}.
A summary and some concluding remarks are given in Sec.~\ref{Sec:Conclusion}.
The source code for simulations in \textsf{RUST} and \textsf{PYTHON 3} are given in the Ancillary Files \cite{supplement}.

%%%%%%%%%%%%%%%%%%%%%%%%%%%%%%%%%%%%%%%%%%%%%%%%%%%%%%%%%%%%%%%%%%%%%%%%%%%
%%% Photocounting probabilities with realistic photon-number resolution
%%%%%%%%%%%%%%%%%%%%%%%%%%%%%%%%%%%%%%%%%%%%%%%%%%%%%%%%%%%%%%%%%%%%%%%%%%%

\section{Photocounting probabilities with realistic photon-number resolution}
\label{Sec:Ph-count_Pr}

In this section we derive a formula for the photocounting distribution in GBS with an arbitrary type of detectors.
First, let us consider the standard GBS scheme.
Gaussian states with no coherent displacement  (including the vacuum states) are injected at the inputs of a linear interferometer.
The modes are analyzed by photocounters at the output of this interferometer.
Each measurement event is represented by a click pattern $\boldsymbol{n}=(n_1,\ldots,n_M)$, where $M$ is the number of the interferometer outputs.
As shown in Ref.~\cite{Hamilton2017}, the probability distribution of this pattern reads
    \begin{align}\label{id_st}
        P^\textrm{(i)}(\boldsymbol{n}) = \frac{1}{\sqrt{|\sigma_Q|}}
        &\prod^M_{i=1}\frac{1}{n_i!}\left( \frac{\partial^2}{\partial \alpha^\ast_i \partial \alpha_i} \right)^{n_i}\\
        &\times\exp{\left(\frac{1}{2}\boldsymbol{\xi}^{\dagger} A \boldsymbol{\xi}\right)}
        \Bigg|_{\boldsymbol{\xi}=0}.\nonumber
    \end{align}
Here $\boldsymbol{\xi}=(\alpha_1, \ldots, \alpha_M, \alpha_1^{*},\ldots,\alpha_M^{*})^\mathrm{T}$ is the vector of complex amplitudes and their complex conjugations, $A=\mathrm{I}-\sigma_Q^{-1}$, and $\sigma_Q$ is the covariance matrix of the $Q$ function for the state at the interferometer outputs, related to the covariance matrix $\sigma$ of the Wigner function as $\sigma_Q=\sigma+\mathbb{I}/2$.
See Ref.~\cite{Hamilton2024} for a recent result generalizing Eq.~(\ref{id_st}) to the case of non-Gaussian states.

The photocounting distribution for detectors with realistic photon-number resolution can be obtained from the general form of the photocounting formula, which in the considered case is given by
    \begin{align}\label{Eq:PhotoCountFormula}
        P(\boldsymbol{n})=\Tr\left[\hat{\Pi}(\boldsymbol{n})\hat{\rho}\right].
    \end{align}
Here $\hat{\rho}$ is the density operator,
    \begin{align}\label{Eq:POVM-n}
        \hat{\Pi}(\boldsymbol{n})=\bigotimes_{i=1}^{M}\hat{\Pi}_{n_i},
    \end{align}
and $\hat{\Pi}_{n_i}$ is the POVM for the detection process of a single mode.
We will use two representations for the POVM.
The first is the Fock-state representation,
    \begin{align}\label{Eq:FockStateRepr}
        P_{n_i|m_j}=\bra{m_j}\hat{\Pi}_{n_i}\ket{m_j}.
    \end{align}
This expression can be interpreted as the probability distribution to get $n_i$ clicks of the detector given $m_j$ photons at its input.
Here $\ket{m_j}$ is the Fock state.
Since the POVM elements $\hat{\Pi}_{n_i}$ for all photocounting techniques commute with the photon-number operator, all off-diagonal terms in Eq.~(\ref{Eq:FockStateRepr}) vanish.
Another representation is given by the $Q$ symbols of the POVM,
    \begin{align}
        \Pi_{n_i}(\alpha^\ast,\alpha)=\bra{\alpha}\hat{\Pi}_{n_i}\ket{\alpha},
    \end{align}
where $\ket{\alpha}$ is a coherent state.
Two representations are related to each other as 
    \begin{align}\label{Pi_P}
        \Pi_{n_i}(\alpha^\ast,\alpha)=
        \sum\limits_{m_j=0}^{\infty}\frac{|\alpha|^{2m_j}}{m_j!}
        e^{-|\alpha|^{2}}
        P_{{n_i}|m_j}.
    \end{align}
Importantly, if the detection process does not involve dark counts, afterpulses, and other clicks that are not directly related to the detected photons, then $P_{n_i|m_j}=0$ for $m_j<n_i$.

Similar to the case of BS, which involves detectors with realistic photon-number resolution \cite{Len2022}, the probability distribution for GBS can be expressed as
    \begin{align}\label{real_ph}
        P(\boldsymbol{n})=
        \sum\limits_{m_{1}=0}^{\infty}\ldots\sum\limits_{m_{M}=0}^{\infty}
        P_{n_1|m_1}\ldots P_{n_M|m_M}
        P^{\textrm{(i)}}(\boldsymbol{m}).
    \end{align}
Substituting Eq.~(\ref{id_st}) into Eq.~(\ref{real_ph}) and taking into account Eq.~(\ref{Pi_P}), we can derive a general expression for the probability distribution for the click pattern $\boldsymbol{n}$ in the case of GBS with an arbitrary photodetector,
    \begin{align}\label{gen_pr}
        P(\boldsymbol{n}) = 
        \frac{1}{\sqrt{|\sigma_Q|}}
        \mathcal{F}^\Pi_{\boldsymbol{n}}\left[A\right].
    \end{align}
Here
    \begin{align}\label{mat_f}
        \mathcal{F}^\Pi_{\boldsymbol{n}}\left[A\right]=
        \prod^M_{i=1} 
        \Pi_{n_i}
        \left( \frac{\partial}{\partial \alpha_i^\ast},
        \frac{\partial}{\partial \alpha_i}
        \right)
        & \exp{\left( \frac{\partial^2}{\partial \alpha_i^\ast \partial \alpha_i} \right)}
        \\\nonumber
        \times
        & \left. \exp{\left(\frac{1}{2}\boldsymbol{\xi}^{\dagger} A \boldsymbol{\xi}\right)}
        \right|_{\boldsymbol{\xi=0}}
    \end{align}
is a functional of the matrix $A$, whose form depends on the POVM for the given detection scheme.
The upper index $\Pi$ in this functional indicates the type of detection.
Examples of the POVMs for common detection techniques are given in Appendix~\ref{App:DetMeth}.

The matrix functional $\mathcal{F}^\Pi_{\boldsymbol{n}}\left[A\right]$ reduces to the already known forms for GBS with the detection methods considered in the literature.
First, let us consider the original variant of GBS \cite{Hamilton2017} with the PNR detectors.
The corresponding POVM is given by Eq.~(\ref{id_POVM}) assuming $\eta=1$.
In this case, the functional $\mathcal{F}^{\Pi}_{\boldsymbol{n}}\left[A\right]$ is expressed in terms of the Hafnian as
    \begin{align}\label{id_haf}
        \mathcal{F}^{\mathrm{PNR}}_{\boldsymbol{n}}\left[A\right]=\frac{1}{\boldsymbol{n}!}
        \Haf\left[XA_{\boldsymbol{n}}\right].
    \end{align}
Here $\boldsymbol{n}!=\prod_{i=1}^{M}n_i!$, $X=\begin{pmatrix}0 & \mathbb{I}\\ \mathbb{I} & 0\end{pmatrix}$,
$\mathbb{I}$ is the $n\times n$ identity matrix and $n=\sum_{i=1}^{M}n_i$ is the total number of clicks.
The $2n\times 2n$ matrix $A_{\boldsymbol{n}}$ is derived from matrix $A$ by retaining solely the rows and columns associated with triggered detectors. 
Each of these selected rows and columns is repeated until their number matches the number of photons at the corresponding output.

Second, let us consider GBS with on-off detectors \cite{Quesada2018}.
In this case, the POVM is given by Eq.~(\ref{Eq:Array}) with $K=1$.
The functional $\mathcal{F}^\Pi_{\boldsymbol{n}}\left[A\right]$ is then reduced to
    \begin{align}\label{Eq:on/off_Tor}
        \mathcal{F}^{\mathrm{on-off}}_{\boldsymbol{n}}\left[A\right]=\Tor\left[A_{S(\boldsymbol{n})}\right],
    \end{align}
where $\Tor A$ is the Torontonian of the matrix $A$ and the matrix $A_{S(\boldsymbol{n})}$ is defined similarly to the matrix $A_{\boldsymbol{n}}$ in the previous case but without repeating rows and columns.
In a more general case of click detectors, the POVM is again given by Eq.~(\ref{Eq:Array}) but with $K\geq 1$.
For such a scenario, experimentally implemented in Ref.~\cite{deng2023}, the above functional is given by
    \begin{align}\label{Eq:array}
    \mathcal{F}^{\mathrm{click}}_{\boldsymbol{n}}\left[A\right]=\Ken\left[A_{S(\boldsymbol{n})}\right].
    \end{align}
Here the matrix functional, the Kensingtonian (cf. Ref.~\cite{bressanini2023}), reads
    \begin{align}\label{Eq:KenTor}
        &\Ken\left[A_{S(\boldsymbol{n})}\right] = 
        \prod_{i=1}^{N} \binom{K}{c_{i}}\\
        &\times\sum\limits_{k_{1}=0}^{c_{1}-1} 
        \dots
        \sum\limits_{k_{N}=0}^{c_{N}-1}
        \prod_{i=1}^{N} \binom{c_{i}}{k_{i}}
        (-1)^{k_{i}}
        \mathrm{Tor}(B_{S(\boldsymbol{n})}),\nonumber
    \end{align}
where 
    \begin{align}\label{Eq:BSmatrix}
        B_{S(\boldsymbol{n})}=\diag\left(\sqrt{\frac{c_i-k_i}{K}}\right) A_{S(\boldsymbol{n})} \diag\left(\sqrt{\frac{c_i-k_i}{K}}\right).
    \end{align}
In these expressions, the numbers $c_i$ are $c_i=n_{l_i}\neq 0$.
That is, they are nonzero numbers of clicks.
The indices $l_i$ belong to the set $S(\boldsymbol{n})=\{l_1,\dots, l_N\}$, and $N$ is the total number of triggered detectors.
Here and in the following the elements of the $2N\times 2N$ matrix $\diag (b_i)$ are defined as $[\diag (b_i)]_{k,l}=[\diag (b_i)]_{k+N,l+N}=b_k\delta_{k,l}$ for $k,l=1,\ldots,N$.
For details see Sec.~\ref{App:ProbabDistr_Click}.

Let us consider another scenario: counting pulses of photocurrent within a measurement-time-window duration of $\tau_\textrm{m}$.
Every pulse corresponds to an absorbed photon.
After a photon is registered, the next one cannot be registered during the detector dead time $\tau_{\mathrm{d}}$.
In the most general case, which is inherent to the SNSPDs, the ability of the detector to register the next photon is smoothly recovered during the relaxation time $\tau_{\mathrm{r}}$.
The most general scenario also assumes that the normalized intensity shape $I(t)$ within each time window may not be rectangular---i.e., the light mode may be significantly nonmonochromatic.
This leads to an inhomogeneous probability distribution of the appearance of the photocurrent pulse in the time instance $t$ inside the measurement time window.

In the most general scenario, inherent in the SNSPDs, the POVM is given by Eq.~(\ref{Eq:SNSPD}).
Its direct application to Eq.~(\ref{mat_f}) gives the expression (see Sec.~\ref{App:ProbabDistr_SNSPDs})
    \begin{align}\label{Eq:FunctPulses}
        &\mathcal{F}^{\mathrm{SNSPD}}_{\boldsymbol{n}}\left[A\right]=
        \int_{T_{c_{1}}} d^{c_{1}} 
        \boldsymbol{t}_{1}
        \ldots
        \int_{T_{c_{N}}} d^{c_{N}} 
        \boldsymbol{t}_{N}
        \prod\limits_{j=1}^{N}
        \mathcal{I}_{c_{j}} (\boldsymbol{t}_{j})
        \\\nonumber
        &\times\frac{\Haf\left(X 
        \Omega_{S(\boldsymbol{n}),\boldsymbol{n}}
        \right)}
        {\sqrt{|\mathbb{I} -\diag\sqrt{1 - \Xi_{c_{i}}(\boldsymbol{t}_{i})} A_{S(\boldsymbol{n})}\diag\sqrt{1 - \Xi_{c_{i}}(\boldsymbol{t}_{i})}|}},
    \end{align}
where $\mathcal{I}_{n} (\boldsymbol{t})$ and $\Xi_{n}(\boldsymbol{t}_{j})$ are given by Eqs.~(\ref{Eq:IntMultiTime}) and (\ref{Eq:MultiTimeDepEff}), respectively.
The $2n\times 2n$ matrix $\Omega_{S(\boldsymbol{n}),\boldsymbol{n}}$ is derived from the matrix
     \begin{align}
        \Omega_{S(\boldsymbol{n})} &=
        \diag(\Xi_{c_{i}}(\boldsymbol{t}_{i}))
        \\\nonumber
        &-\left((1-A_{S(\boldsymbol{n})})^{-1}
        +\diag\left(\frac{1}{ \Xi_{c_{i}}(\boldsymbol{t}_{i})}-1\right) \right)^{-1}
    \end{align}
by repeating each $i$th row and column until their number matches the corresponding number of clicks (pulses) $c_i$.

For the case of APDs, $\tau_\textrm{r}=0$, and the rectangular intensity shape [see Eq.~(\ref{Eq:RecEnv})], the integrals in Eq.~(\ref{Eq:FunctPulses}) can be evaluated analytically for $n_i=\{0,1\}$ (see Sec.~\ref{App:ProbabDistr_APDS}); i.e., for the collision-free subspace,
\begin{align}\label{Eq:FAPD}
    \mathcal{F}^{\mathrm{APD}}_{\boldsymbol{n}}\left[A\right]
    =\sum\limits_{Z \in P[N]}
    \frac{(-1)^{|Z|} \Haf(D_{S(\boldsymbol{n}),Z})}
    {\sqrt{\left|\mathbb{I} 
    - (1-\eta_1) A_{S(\boldsymbol{n}),Z}
    \right|}},
\end{align}
where
    \begin{align}\label{Eq:D_S}
        &D_{S(\boldsymbol{n}),Z}=X_{|Z|}
        \\\nonumber
        &\times
        \left\{2\mathbb{I} -
        \left[\eta_1(\mathbb{I} - A_{S(\boldsymbol{n}),Z})^{-1} 
        + (1-\eta_1)\mathbb{I} \right]^{-1}\right\}.
    \end{align}
Here $P([N])$ is the power set (the set of all subsets) of $[N]:=\{1,2,\ldots,N\}$, $Z$ denotes its elements, and $|Z|$ is the number of these elements, i.e., the cardinality. 
The $2|Z|\times 2|Z|$ matrix $A_{S(\boldsymbol{n}),Z}$ is obtained from the matrix $A_{S(\boldsymbol{n})}$ by retaining solely the rows and columns related to each element of the set $Z$.

%%%%%%%%%%%%%%%%%%%%%%%%%%%%%%%%%%%%%%%%%%%%%%%%%%%%%%%%%%%%%%%%%%%%%%%%%%%
%%% Validation of GBS with realistic photon-number resolution
%%%%%%%%%%%%%%%%%%%%%%%%%%%%%%%%%%%%%%%%%%%%%%%%%%%%%%%%%%%%%%%%%%%%%%%%%%%

\section{Validation of GBS with realistic photon-number resolution}
\label{Sec:Orbit_pr}

As mentioned in the Introduction, full certification is infeasible in the GBS scenario.
Obviously, any sampling for detectors with realistic photon-number resolution can be used to derive a sampling for on-off detectors.
Since the latter can be computationally hard (cf. Ref.~\cite{Quesada2018}), the same can be stated for the former. 
Therefore, for validation purposes we need to consider marginal or grouped probabilities that can be reconstructed in experiments and compare them with classically modeled distributions. 
An example of such coarse-grained events is given by orbits considered in Refs.~\cite{Bradler2021,Schuld2020,Giordani2023}.  

In the most general case, the orbit $\mathcal{O}_{[n_1,\ldots,n_M]}$ is a group of click patterns that can be obtained from a given pattern $\mathbf{n}=\{n_1,\ldots,n_M\}$ by permutations of its components.  
The probability of the orbit can be obtained as the sum of the probabilities for each click pattern.
Typically, such a coarse-grained probability can be estimated from the sampling data, while the same is impossible for the probability of an individual click pattern.
In the case of ideal PNR detectors, the probabilities of orbits are related to the feature vectors of graphs encoded in the device (see, e.g., \cite{Schuld2020,Arrazola2018a,Arrazola2018b}).
In the GBS scenarios with realistic GBS, such a connection with graph theory is generally unclear.

Following the idea of Ref.~\cite{Giordani2023}, we will use orbits with almost all collision-free events and a few events with two clicks.
For such orbits we will use the notation $\mathcal{O}_l^{n}$, where $l$ is the number of outputs with two detected clicks and $n$ is the total number of clicks.
We have tailored the two methods presented in Ref.~\cite{Schuld2020} and in Refs.~\cite{Opanchuk2018,Drummond2022} to detectors with realistic photon-number resolution to estimate the probability of such orbits on a classical device.

\subsection{Orbit probability estimation}
\label{SubSec:Orb_est}

The first method to evaluate the probability of orbits (cf. Ref.~\cite{Schuld2020}) consists of the following steps.
First, we randomly select $N_S$ click patterns $\mathbf{n}_i\in\mathcal{O}_l^n$.
Second, we compute the corresponding probabilities $P (\boldsymbol{n}_i)$.
Notably, we still have a way to compute the Torontonian \cite{kaposi2022} and Hafnian for a small number of photons, in our case up to $n=16$.
Finally, we approximate the orbit probability as
    \begin{align}\label{apr_orb}
        \mathcal{P}(\mathcal{O}_l^n) \approx 
        \frac{|\mathcal{O}_l^n|}{N_{S}} \sum\limits_{i=1}^{N_S} P (\boldsymbol{n}_i),
    \end{align}
where $|\mathcal{O}_l^n|$ is the number of click patterns in the orbit.
The accuracy of such a procedure was considered in Ref.~\cite{Schuld2020} based on a result presented in Ref.~\cite{Shervashidze2009}.
This method is efficiently used to estimate the orbit probabilities in the case of ideal PNR, on-off, and click detectors for small $n$.
We use it to control our results obtained with the second method.

The second method \cite{Opanchuk2018,Drummond2022} is based on the technique of phase-space simulation from the positive $P$ function \cite{Drummond1980, Drummond1981}.
We have tailored this method to estimate the orbit probabilities $\mathcal{P}(\mathcal{O}_l^n)$ in the case of GBS with realistic photon-number resolution.
In contrast to the first method, this simulation technique shows high scalability and computational speed, making it applicable to larger numbers of photons. 

Let $m_1$ and $m_2$ be a number of outputs with one and two clicks, respectively, i.e., $l=m_2$ and $n=m_1+2m_2$.
This means that $\mathcal{O}_l^{n}=\mathcal{O}_{m_2}^{m_1+2m_2}$.
Following Eq.~(\ref{Eq:PhotoCountFormula}), the orbit probability in this case is given by
    \begin{align}\label{Eq:OrbitProb}
        \mathcal{P}(\mathcal{O}^{m_1+2 m_2}_{m_2})=
        \Tr\left(\hat{\rho} \sum\limits_{\boldsymbol{n} \in 
        \mathcal{O}^{m_1+2 m_2}_{m_2}} 
        \hat{\Pi}(\boldsymbol{n})
        \right).
    \end{align}
Formally, we also assume that $\mathcal{P}(\mathcal{O}^{m_1+2 m_2}_{m_2})=0$ for $m_1+m_2>M$.
Next, we consider the discrete characteristic function of this un-normalized probability distribution of orbits by providing the corresponding discrete Fourier transform with respect to the variables $m_1$ and $m_2$,
    \begin{align}\label{Eq:CharF1}
        C\left(k_1,k_2\right)=\sum\limits_{m_1,m_2=0}^M\mathcal{P}(\mathcal{O}^{m_1+2 m_2}_{m_2})
        e^{-i (k_1 m_1+k_2 m_2)\theta}.
    \end{align}
Combining this expression with Eqs.~(\ref{Eq:POVM-n}) and (\ref{Eq:OrbitProb}), we get
    \begin{align}\label{Eq:CharF2}
        C&\left(k_1,k_2\right)\\
        &=\Tr\left[\hat{\rho}\left(
        \hat{\Pi}_{0}+\hat{\Pi}_{1} e^{-i k_1 \theta} + \hat{\Pi}_{2} e^{-i k_2 \theta}\right)^{\otimes M}
        \right],\nonumber
    \end{align}
where $\theta = \frac{2\pi}{M + 1}$ and $k_{1,2} = 0,\dots, M$. 

Equation~(\ref{Eq:CharF2}) can be rewritten in the generalized $P$ representation as
    \begin{align}\label{Eq:CharF3}
        C\left(k_1,k_2\right)&=
         \Bigg\langle\prod_{i=1}^M \big[\pi_i(0|\alpha_i,\beta_i)\\
         &+\pi_i(1|\alpha_i,\beta_i) e^{-i k_1 \theta}
         + \pi_i(2|\alpha_i,\beta_i) e^{-i k_2 \theta} \big]\Bigg\rangle_P.\nonumber
    \end{align}
Here
\begin{align}\label{Eq:POVM_Symb}
    \pi_i(q|\alpha_i,\beta_i)= \bra{\beta_{i}^{*}}
    \hat{\Pi}_{q}
    \ket{\alpha_{i}},
\end{align}
and $\ket{\alpha}$ is the coherent state.
The averaging is taken over the positive $P$ function $P(\boldsymbol{\alpha},\boldsymbol{\beta})$, where $\boldsymbol{\alpha}=(\alpha_1,\ldots,\alpha_M)^{\mathrm{T}}$ and $\boldsymbol{\beta}=(\beta_1,\ldots,\beta_M)^{\mathrm{T}}$.
Based on this expression, the method can be summarized as follows: (1) sampling complex amplitudes $\boldsymbol{\alpha}$ and $\boldsymbol{\beta}$ from the positive $P$ function (see Refs.~\cite{Opanchuk2018,Drummond2022} and Appendix~\ref{App:PhaseSpaceSimm} for details), (2) estimating the characteristic function~(\ref{Eq:CharF3}) from the generated sample set, and (3) using the inverse discrete Fourier transform,
    \begin{align}\label{Eq:DIFT}
        \mathcal{P}&\left(\mathcal{O}^{m_1+2 m_2}_{m_2}\right) =\frac{1}{\left(M+1\right)^2} 
        \\\nonumber
        &\times\sum\limits_{k_1=0}^M \sum\limits_{k_2=0}^M C(k_1, k_2) 
        e^{i (k_1 m_1 + k_2 m_2) \theta},
    \end{align}
to reconstruct the probabilities of the orbits $\mathcal{P}\left(\mathcal{O}^{m_1+2 m_2}_{m_2}\right)$.

\subsection{Validation procedure}

To demonstrate the role of realistic photon-number resolution for validation procedures, we provide simulations of orbit probabilities for two cases, labeled A and B (see Table~\ref{Tab:Cases}).
In both cases, thermalized squeezed vacuum states are supposed to be injected into $M^\prime$ input ports of an interferometer with $M$ output ports.
These states are characterized by the covariance matrix
   	\begin{align}\label{Eq:SVS}
    		\sigma=\frac{1}{2}\begin{pmatrix}
    		    \cosh2r & (1-\epsilon)\sinh2r\\
                (1-\epsilon)\sinh2r & \cosh2r
    		\end{pmatrix},	
    \end{align}
where $r$ is the squeezing parameter and $\epsilon=0.1$ is the thermalization factor.
The overall efficiency, which includes both interferometer and detector losses, is $\eta=0.8$.
The squeezing parameter is chosen such that the expectation value of the total number of photons in all ports after all losses is $n_{\mathrm{ph}}$.

\begin{table}[h!]
    \centering
     \caption{\label{Tab:Cases} Parameters of the GBS devices for cases A and B used for simulations in this paper.
     Here $M$ is the number of output ports, $M^\prime$ is the number of input ports with injected squeezed states, $r$ is the squeezing parameter, and $n_{\mathrm{ph}}$ is the mean number of photons in all ports after all losses.}
     \setlength{\tabcolsep}{12.0pt}
    \begin{tabular}{ccccc}
    \hline\hline
         Case & $M$ & $M^\prime$ & $r$ & $n_{\mathrm{ph}}$ \\
         \hline\hline
        A & 400 & 200 & 0.3466 & 20 \\
        \hline
        B & 144 & 50 & 1 & 55.24\\
        \hline\hline
    \end{tabular}
\end{table}

\begin{figure}[ht!]
                \includegraphics[width=1\linewidth]{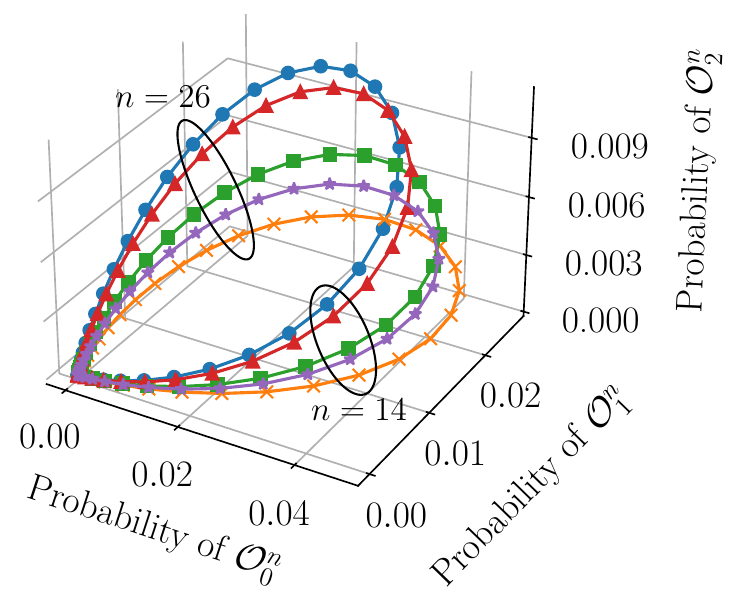}
                \caption{The probabilities for the orbits $\{\mathcal{O}_0^n, \mathcal{O}_1^n, \mathcal{O}_2^n\}$ in case A (see Table~\ref{Tab:Cases}) for different numbers of clicks $n$ are shown for thermalized squeezed vacuum states at input ports.
                See the text for parameters characterizing the states, losses, and interferometer.
                Blue circles correspond to the ideal PNR detector.
                Red triangles correspond to counting pulses within a measurement-time window with an APD and $\tau_\mathrm{d}=0.05\tau_\mathrm{m}$.
                Purple stars correspond to the same technique and dead time with SNSPDs with relaxation time $\tau_\mathrm{r}=0.2\tau_\mathrm{m}$.
                Orange crosses and green squares correspond to click detectors with $K=2$ and $K=3$ on-off detectors in the array, respectively.}
                \label{Fig:pd_ad2_ad3_dtd_nw}
        \end{figure}

The probabilities of orbits $\mathcal{O}_l^n$ for various detection techniques in case A are shown in Fig.~\ref{Fig:pd_ad2_ad3_dtd_nw} for different click numbers $n$.
As intuitively expected, the probabilities of the orbits $\mathcal{O}_2^n$ are highest for the ideal PNR detectors.
They are significantly smaller for click detectors with $K=2$ and $K=3$ on-off detectors in the array.
We also consider counting photocurrent pulses for an APD within a measurement-time window of duration $\tau_\textrm{m}$ with dead time $\tau_\mathrm{d}=0.05\tau_\mathrm{m}$.
This results in a curve that is relatively close to the curve associated with ideal PNR detectors.
However, when using an SNSPD with the same dead time and relaxation time $\tau_\textrm{r}=0.2\tau_\textrm{m}$, the corresponding curve is characterized by significantly smaller probabilities of orbits $\mathcal{O}_2^n$.
Therefore, for validation techniques based on estimations of the probabilities $\mathcal{P}\left(\mathcal{O}_l^n\right)$, it is crucial to consider the effect of imperfect photon-number resolution.

    \begin{figure*}[ht!]
        \centering
        \center{\includegraphics[width=1\linewidth]{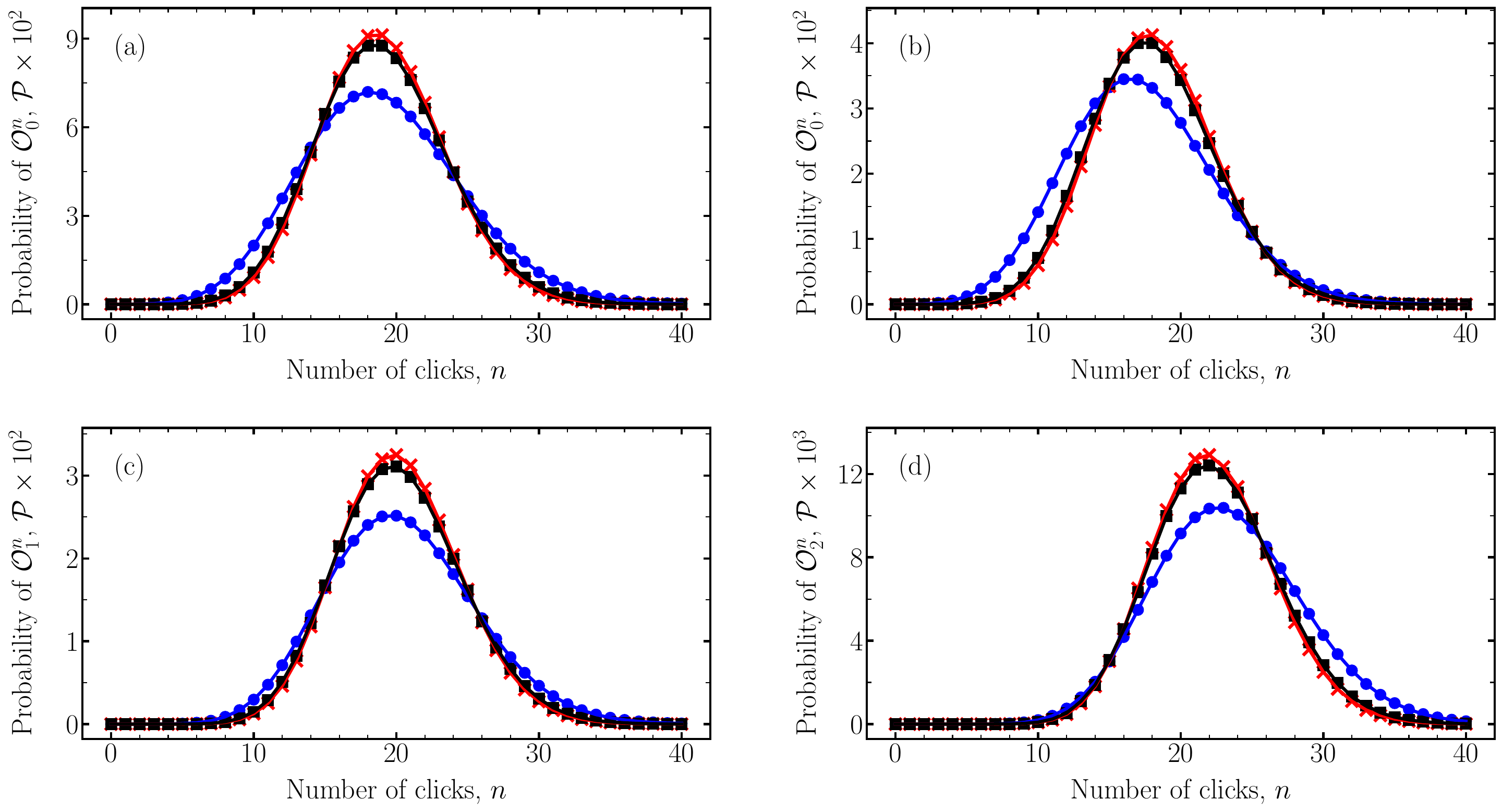}}
        \caption{The probabilities for orbit set $\{\mathcal{O}_0^n, \mathcal{O}_1^n, \mathcal{O}_2^n\}$ for different numbers of clicks $n$ in case A (see Table~\ref{Tab:Cases}) are shown.
        (a) corresponds to on-off detectors.
        (b)-(d) correspond to counting photocurrent pulses within a measurement time window with an APD and dead time $\tau_{\textrm{d}}=0.05\tau_{\textrm{m}}$.
        Blue circles, red crosses, and black squares correspond to the thermalized squeezed vacuum state, thermal states, and squashed states, respectively.
        See the text for parameters characterizing the states, losses, and interferometer.}
        
        \label{fig:DTD_orb}
    \end{figure*}

For validation, we propose to compare the orbit probabilities $\mathcal{P}\left(\mathcal{O}_l^n\right)$ for the thermalized squeezed vacuum states with the same probabilities obtained for classical states characterized by non-negative $P$ functions.
For the latter states, we provide direct classical simulations of the click patterns $\mathbf{n}$ using the method of Ref.~\cite{Rahimi-Keshari2016}.
In particular, we use inputs with thermal and squashed states, characterized by the covariance matrices
   	\begin{align}\label{Eq:Thermal}
    \sigma=\frac{1}{2}\begin{pmatrix}
    		1+2n_{\mathrm{th}} & 0\\
                0 & 1+2n_{\mathrm{th}}
    		\end{pmatrix}
      \end{align}
and
  \begin{align}
      \sigma=\frac{1}{2}
        \begin{pmatrix}
    		1+2n_{\mathrm{th}} & 2n_{\mathrm{th}}\\
                2n_{\mathrm{th}} & 1+2n_{\mathrm{th}}
        \end{pmatrix},	
    \end{align}
respectively.
Here $n_{\mathrm{th}}$ is the number of thermal photons, chosen such that the expectation number of photons in all ports after all losses is $n_{\mathrm{ph}}$.

Unlike the method used in Ref.~\cite{Giordani2023}, the phase-space simulation technique enables us to estimate the orbit probabilities $\mathcal{P}\left(\mathcal{O}_l^n\right)$ for large values of $n$.
In Fig.~\ref{fig:DTD_orb} we show the orbit probabilities in case A for thermal, squashed, and thermalized squeezed vacuum states as a function of the number of clicks $n$ for counting photocurrent pulses by an APD within a measurement time window and for the on-off detectors.
The plots for the thermal and squashed states are markedly different from the plots for the thermalized squeezed vacuum states in all cases.
This difference when using squashed states and on-off detectors was shown in Ref.~\cite{dellios2023a}.

\subsection{Pearson's $\chi^2$ test}

\begin{table*}
\caption{\label{Tab:Comparisson}
The chi-square statistics $\chi^2/k_l$ for different types of detection techniques in case A (see Table~\ref{Tab:Cases}) are given.
Parameters for counting pulses with APDs and SNSPDs are the same as in Fig.~\ref{Fig:pd_ad2_ad3_dtd_nw}.}
\begin{center}
\setlength{\tabcolsep}{12.0pt}
\begin{tabular} {l c c c c c c}
	\hline
	\hline
	\multirow{ 2}{*}{Detector or detection technique}& \multicolumn{3}{ c }{Thermal states} & \multicolumn{3}{c}{Squashed states} \\
	& $\mathcal{O}^n_{0}$ & $\mathcal{O}^n_{1}$ & $\mathcal{O}^n_{2}$ & $\mathcal{O}^n_{0}$ & $\mathcal{O}^n_{1}$ & $\mathcal{O}^n_{2}$ \\
	\hline
	\hline
	Ideal PNR detector &  $1.1\times 10^4$ &$5.3\times 10^3$&$3.1\times 10^3$&  $7.9\times 10^3$&$3.6\times 10^3$&$2.1\times 10^3$
	\\
    \hline
	On-off detector  & $2.0\times 10^4$&     &      & $1.5\times 10^4$&    &   
	\\\hline
	Click detector, $K=2$  & $1.4\times 10^4$&$5.5\times 10^3$&$2.4\times 10^3$ &  $9.8\times 10^3$&$3.8\times 10^3$&$1.6\times 10^3$
	\\\hline
	Click detector, $K=3$  & $1.3\times 10^4$&$5.5\times 10^3$&$2.8\times 10^3$ &  $9.0\times 10^3$ &$3.8\times 10^3$&$1.9\times 10^3$
	\\\hline
	Counting pulses with APD &  $1.2\times 10^4$&$5.3\times 10^3$ &$3.2\times 10^3$ &  $8.4\times 10^3$&$3.7\times 10^3$&$2.1\times 10^3$
	\\\hline
	Counting pulses with SNSPDs &   $1.7\times 10^2$& $6.3\times 10^1$&  $3.4\times 10^1$ &   $1.2\times 10^2$& $4.3\times 10^1$& $1.9\times 10^1$ 
	\\\hline\hline
\end{tabular}
\end{center}
\end{table*}

In order to quantitatively characterize the difference between classical and quantum statistics, we use the Pearson $\chi^2$ method, similar to what was done in Ref.~\cite{Drummond2022} for grouped probabilities.
For this purpose, we generate a set of click patterns for classical states:
$10^{5}$ patterns for counting photocurrent pulses with the SNSPDs and $10^{7}$ patterns for other detection techniques. 
Then we select only those orbits for which the corresponding patterns appear more than 10 times.
Based on these data, we estimate the probabilities of different orbits for classical states, $\mathcal{P}^{\textrm{cl}}\left(\mathcal{O}^n_l\right)$.
For the quantum statistics and the estimated classical statistics we calculate the conditional (normalized) probability distribution of the total number of clicks $n$ given the number of outputs with two clicks $l$,
    \begin{align}
        \mathcal{P}(n|l)=\frac{\mathcal{P}\left(\mathcal{O}^n_l\right)}{\sum_{n^\prime} \mathcal{P}\left(\mathcal{O}^{n^\prime}_l\right)}.
    \end{align}
Let us number the orbits with a given $l$ by the index $i=1,\ldots,k_l$ and denote the total number of clicks in the $i$th orbit as $n[i]$.
Here $k_l$ is the number of considered orbits $\mathcal{O}^{n[i]}_l$ for the given $l$ and for all $i$.
The $\chi^2$ statistics is evaluated as
    \begin{align}
        \chi^2(l)=
        N_l    
        \sum\limits_{i=1}^{k_l} \frac{\left\{\mathcal{P}\big(n[i]|l\big)-\mathcal{P}^{\textrm{cl}}\big(n[i]|l\big)\right\}^2}{\mathcal{P}\big(n[i]|l\big)},
    \end{align}
where $N_l$ is the number of considered and selected click patterns for which the number of outputs with two clicks is equal to $l$.

Results from similar statistics should have $\chi^2(l)/k_{l}\approx 1$. 
In the case of significant discrepancy, $\chi^2(l)/k_{l} \gg 1$ should hold. 
The test results for various detection techniques in case A are presented in Table~\ref{Tab:Comparisson}.
From these data we can conclude that the presented $\chi^2$ test enables us to distinguish between classical and quantum statistics with high confidence.
It is worth noting that the large values of $\chi^2/k_l$ for different cases depend on a variety of factors, such as the numbers $N_l$ and $k_l$, the size of the sample sets, the types of detection techniques, etc.
Therefore, these values cannot be used directly to compare the degree of discrepancy between different cases.
We also note that although the scenario with on-off detectors also gives large values of  $\chi^2/k_0$, the (im)perfect ability of detectors to discriminate between the numbers of photons gives additional confidence in the discrimination between classical and quantum statistics.

\subsection{Bayesian test}

\begin{figure}[ht!]
    \centering
    \includegraphics[width=0.95\linewidth]{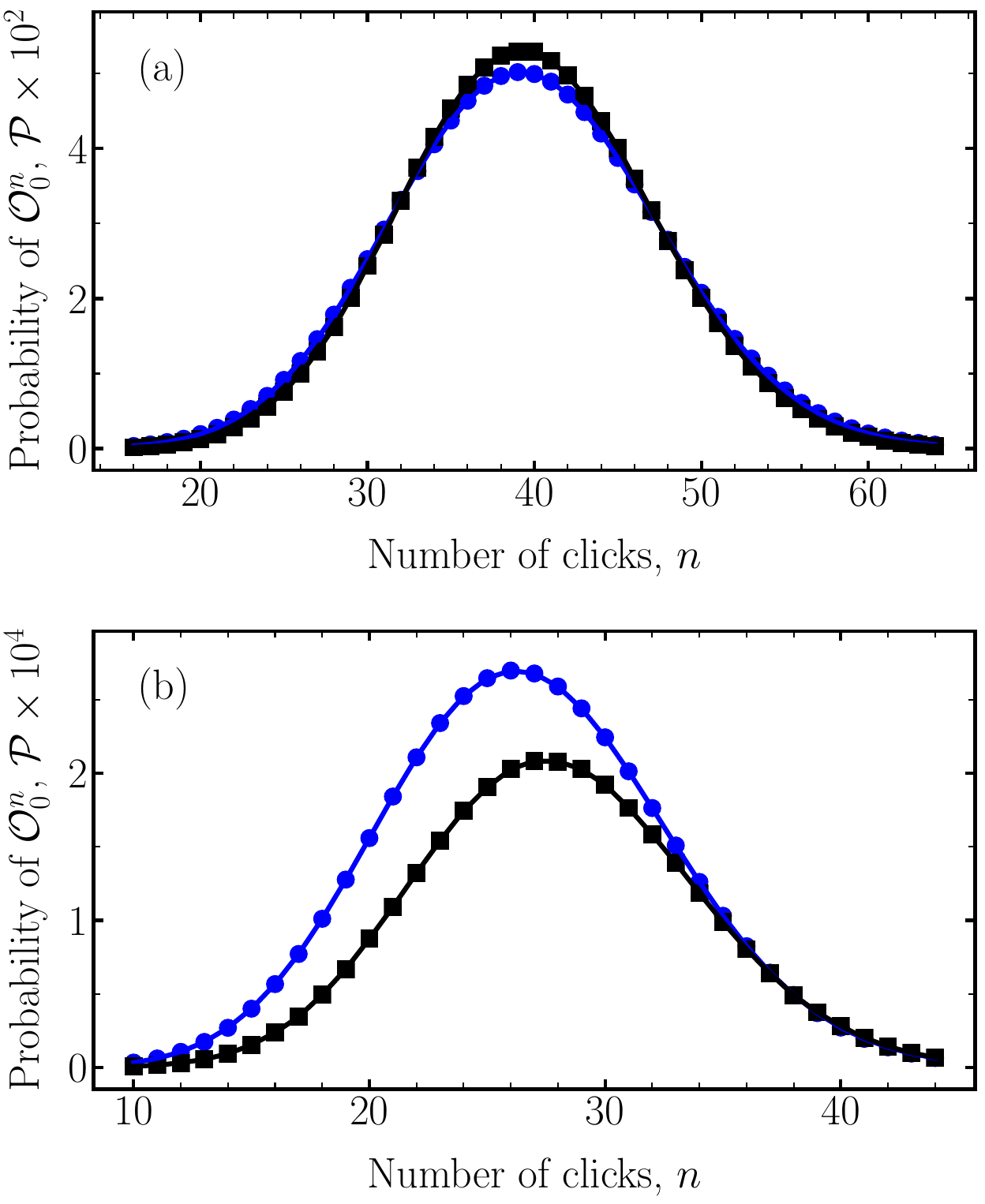}
    \caption{\label{Fig:DTD4} The probabilities for the orbits $\mathcal{O}_0^n$ vs the click number $n$ in case B (see Table~\ref{Tab:Cases}) are shown. 
    (a) corresponds to on-off detectors.
    (b) corresponds to counting photocurrent pulses with an APD, such that $\tau_{\textrm{d}}=0.25\tau_{\textrm{m}}$.
    Blue circles and black squares correspond to the thermalized squeezed vacuum states and the squashed states (estimated with $10^9$ sample events), respectively.}
\end{figure}

Let us consider case B from Table~\ref{Tab:Cases}.
Although the statistics in the scenario with squashed states for $10^7$ sample events in the case of on-off detectors are distinguished from the scenario with thermalized squeezed states with $\chi/k_0=9\times 10^2$, qualitatively, this difference does not have high confidence (see Fig.~\ref{Fig:DTD4}).
However, this difference more reliable if we consider even minimal photon-number resolution, as can be seen in Fig.~\ref{Fig:DTD4} for the example of the orbit probability $\mathcal{P}(\mathcal{O}_0^n)$ obtained by counting photons with the APD for $\tau_{\mathrm{d}}=0.25\tau_{\mathrm{m}}$.

To characterize the advantages of (im)perfect photon-number resolution quantitatively, we use the Bayesian test \cite{Bentivegna2014}.
Simulations of the click patterns $\boldsymbol{n}$ for a nonclassical input state are computationally hard.
For this reason, we simulate a sampling set $\{n_{(i)},l_{(i)}|i=1,\ldots,N_{\mathrm{O}}\}$ of $N_{\mathrm{O}}$ pairs of numbers $n$ and $l$ related to the orbits with $l=0,1,2$ and non-vanishing probabilities.
Each sample event in this simulation corresponds to the orbit to which the particular pattern $\boldsymbol{n}$ belongs without its specification. 
The simulation is based on the normalized probability distribution
    \begin{align}\label{Eq:OrbitProb_H}
        \mathcal{P}(n,l)=\frac{\mathcal{P}\left(\mathcal{O}^n_l\right)}{\sum_{n^\prime,l^\prime} \mathcal{P}\left(\mathcal{O}^{n^\prime}_{l^\prime}\right)},
    \end{align}
where the probabilities $\mathcal{P}\left(\mathcal{O}^n_l\right)$ are obtained by phase-space simulations.
Like in Eq.~(\ref{Eq:OrbitProb_H}), we define probabilities $\mathcal{P}^{\mathrm{cl}}(n,l)$ for the case of squashed states at the input.
The Bayesian confidence is defined as (see Ref.~\cite{Zhong2021})
    \begin{align}
        \Delta H=\frac{1}{N_{\mathrm{O}}}\sum\limits_{i=1}^{N_{\mathrm{O}}}
        \ln\frac{\mathcal{P}(n_{(i)},l_{(i)})}{\mathcal{P}^{\mathrm{cl}}(n_{(i)},l_{(i)})}.
    \end{align} 
If $\Delta H > 0$, then the hypothesis of thermalized squeezed states at the input is more likely than the hypothesis of squashed states.
Moreover, larger values of $\Delta H$ imply larger deviations of quantum statistics from classical.

The Bayesian confidence $\Delta H$ distinguishing GBS with thermalized squeezed states from the case with squashed states at the input in case B from Table~\ref{Tab:Cases} is given in Table~\ref{Tab:Bayes}.
From these data we observe that the usage of click detectors with small numbers $K$ and APDs with small $\tau_{\mathrm{d}}/\tau_{\mathrm{m}}$ leads to higher values of $\Delta H$ compared to the case of on-off detectors.
Thus, even minimal photon-number resolution can significantly increase the confidence in distinguishing between quantum and classical statistics with Bayesian validation methods.

\begin{table}[ht!]
    \centering
     \caption{\label{Tab:Bayes}The Bayesian confidence $\Delta H$ for distinguishing the orbit statistics of the thermalized squeezed states from the squashed states in case B from Table~\ref{Tab:Cases}. 
     The number of sample events $N_{\mathrm{O}}=10^7$.}   
    \setlength{\tabcolsep}{12.0pt}
    \begin{tabular}{l c}
    \hline\hline
        Detector or detection technique & $\Delta H$  \\
    \hline\hline
        On-off detector & $0.0035$ \\
    \hline
        Click detector, $K=2$ & {$0.013$} \\
    \hline
        Click detector, $K=3$ & {$0.017$} \\
     \hline
        Counting pulses with APD, $\tau_{\mathrm{d}}=\tau_{\mathrm{m}}/3$ & {$0.011$} \\ 
     \hline   
        Counting pulses with APD, $\tau_{\mathrm{d}}=0.25\tau_{\mathrm{m}}$ & {$0.015$} \\
     \hline
        {Ideal PNR detector} & {$0.026$} \\
     \hline\hline
    \end{tabular}
\end{table}

%%%%%%%%%%%%%%%%%%%%%%%%%%%%%%%%%%%%%%%%%%%%%%%%%%%%%%%%%%%%%%%%%%%%%%%%%%%
%%% Conclusion
%%%%%%%%%%%%%%%%%%%%%%%%%%%%%%%%%%%%%%%%%%%%%%%%%%%%%%%%%%%%%%%%%%%%%%%%%%%
    
\section{Conclusion}
\label{Sec:Conclusion}

Registrations of collision events in GBS are characterized by a low probability because of the preservation of classical computational hardness.
Nevertheless, these events are used for validation purposes.
Therefore, the ability of detectors to discriminate between numbers of photons can play a crucial role.
 
We derived a general expression for the photocounting probability distribution at the output of the GBS device, assuming general realistic detectors characterized by POVMs.
The result depends on a matrix functional of the covariance matrix of the output state.
It is reduced to known forms, e.g., Hafnian, Torontonian, and Kensingtonian in the special cases of ideal PNR or on-off detectors.
We extended these sets of functionals to the photocounting techniques, which are based on counting pulses of photocurrent within a measurement time window.
The general equation includes the shapes of the output mode and the time-dependent recovery efficiency of the detectors and generally requires numerical integration.
However, it is reduced to an analytical form for collision-free events in the case of the rectangular mode and use of APDs, which are usually characterized by a negligible relaxation time.

We tailored validation protocols for GBS to the case of detectors with realistic photon-number resolution.
This protocol assumes the estimation of probabilities of orbits---group events that contain click patterns obtained from each other by mutual permutations.
In particular, we focused on three types of orbits characterized by zero, one, and two outputs registering collision events.
The corresponding coarse-grained probabilities can be estimated from experimental data, in contrast to the probabilities of individual click patterns.

The method of phase-space simulations based on the positive $P$ function showed a high applicability to estimate orbit probabilities for a high number of output photons.
This task is computationally hard with other simulation techniques.
We estimated the orbit probabilities for two cases, A and B, in which squeezed states are injected into 200 and 50 input ports of interferometers with 400 and 144 output ports, assuming that the mean number of output photons is 20 and 55.24, respectively.

Our results show that detailed information about the POVM of detectors is crucial for a proper analysis of the output statistics.
In particular, it can be important for validation techniques, for example, when using the Bayesian test in case B.
It is worth noting that even the statistics of collision-free events are modified by the imperfect ability of detectors to distinguish between the number of photons.
In our opinion, this factor should be taken into account in relevant theoretical and experimental studies of GBS.

I.S.Y. and A.A.S. appreciate support from the National Research Foundation of Ukraine through Project No.~2020.02/0111, "Nonclassical and hybrid correlations of quantum systems under realistic conditions."
The authors thank J. Sperling, A. Dellios, and G. Bressanini for useful comments.

%%%%%%%%%%%%%%%%%%%%%%%%%%%%%%%%%%%%%%%%%%%%%%%%%%%%%%%%%%%%%%%%%%%%%%%%%%%
%%% appendix
%%%%%%%%%%%%%%%%%%%%%%%%%%%%%%%%%%%%%%%%%%%%%%%%%%%%%%%%%%%%%%%%%%%%%%%%%%%

\appendix

%%%%%%%%%%%%%%%%%%%%%%%%%%%%%%%%%%%%%%%%%%%%%%%%%%%%%%%%%%%%%%%%%%%%%%%%%%%
%%% POVM for detection techniques with realistic photon-number resolution
%%%%%%%%%%%%%%%%%%%%%%%%%%%%%%%%%%%%%%%%%%%%%%%%%%%%%%%%%%%%%%%%%%%%%%%%%%%

\section{POVMs for detection techniques with realistic photon-number resolution}
\label{App:DetMeth}

In this appendix, we list POVMs for various detection techniques that enable an approximate resolution between photon numbers.
We start with the ideal PNR detectors \cite{kelley64,mandel_book}, the POVM for which is given by
\begin{align}\label{id_POVM}
    \hat{\Pi}_k=\hat{F}_k\left(1\right)=\ket{k}\bra{k},
\end{align}
where $\ket{k}$ is the Fock state,
\begin{align}
    \hat{F}_{k}\left(\eta\right)= 
    :\frac{(\eta \hat{n})^{k}}{k!}
    e^{-\eta \hat{n}}:    
\end{align} 
is the POVM for PNR detectors with losses characterized by the efficiency $\eta\in[0,1]$, $\hat{n}$ is the photon-number operator, and $:\ldots:$ denotes normal ordering. 

The counting technique for click detectors is based on a spatial \cite{paul1996, castelletto2007, schettini2007, blanchet08} or temporal \cite{achilles03, fitch03, rehacek03} splitting of a light mode into $K$ modes and detecting each of them separately with an on-off detector.
The corresponding POVM was derived in \cite{sperling12a},
    \begin{align}\label{Eq:Array}
        \hat{\Pi}_k=
        \binom{K}{k}
        :\left( 1 - e^{-\frac{\hat{n}}{K}}
        \right)^{k}
        e^{-\hat{n}\frac{(K-k)}{K}}:.
    \end{align}
For $K=1$ we get the POVM for the on-off detector.

Consider a detection technique based on counting photocurrent pulses within a measurement time window of duration $\tau_\mathrm{m}$.
In this case, dead time (for APDs and SNSPDs) and relaxation of the detector to a previous state (for SNSPDs) result in missing detection events.
In the general case, the corresponding POVM is given by, cf. Ref.~\cite{Uzunova2022},
\begin{align}\label{Eq:SNSPD}
    \hat{\Pi}_n=: \hat{n}^n \int_{T_n} d^n \mathbf{t} \mathcal{I}_n(\mathbf{t}) \exp \left[-\hat{n} \Xi_n(\mathbf{t})\right]:,
\end{align}
where integration is performed over the time-ordering domain
$T_n$ such that $0 \leq t_1 \leq t_2 \leq \ldots \leq t_n \leq \tau_{\mathrm{m}}$, 
\begin{align}\label{Eq:IntMultiTime}
    \mathcal{I}_n(\mathbf{t})=I(t_1)\prod_{i=2}^{n} 
    I(t_i) \xi\left(t_i-t_{i-1}\right),
\end{align}
\begin{align}\label{Eq:MultiTimeDepEff}
    \Xi_n(\mathbf{t})=\int_0^{t_1} d t I(t)+\sum\limits_{i=1}^{n-1} \int_{t_i}^{t_{i+1}} d t I(t) \xi\left(t-t_i\right) 
    \\\nonumber
    +\int_{t_n}^{\tau_{\mathrm{m}}} d t I(t) \xi\left(t-t_n\right)
\end{align}
In these expressions, $I(t)$ is the normalized intensity shape, and $\xi(t)$ is the time-dependent efficiency, describing the relaxation of the SNSPDs after registering a photon.

For the SNSPDs one could use a model for the time-dependent efficiency
\begin{align}\label{Eq:TimeDepEff}
    \xi(t)=\theta\left(t-\tau_{\mathrm{d}}\right) \eta_{\mathrm{r}}\left(t-\tau_{\mathrm{d}}\right).
\end{align}
Here $\tau_\mathrm{d}$ is the dead time, $\theta\left(t-\tau_{\mathrm{d}}\right)$ is the Heaviside step-function, and $\eta_{\mathrm{r}}(t)$ is the recovering efficiency.
The latter can be modeled as
\begin{align}\label{Eq:RelaxTime}
    \eta_{\mathrm{r}}(t)=1-\exp \left(-\frac{t}{\tau_{\mathrm{r}}}\right),
\end{align}
where $\tau_\mathrm{r}$ is the relaxation time.

Let us consider this detection technique with the APDs.
In this case the relaxation time is negligible, i.e., $\eta_{\textrm{r}}=0$.
We also assume that the quantum state is prepared for a rectangular mode, i.e., 
\begin{align}\label{Eq:RecEnv}
    I(t)=\frac{1}{\tau_{\mathrm{m}}}.
\end{align}
This scenario was considered for classical light in Refs.~\cite{ricciardi66,muller73,muller74,cantor75,teich78,vannucci78,rapp2019} and for quantum light in Ref.~\cite{semenov2023}.
The corresponding POVM reads
\begin{align}\label{Eq:Aval0}
    \hat{\Pi}_0=\hat{F}_0(\eta)
\end{align}
for $n=0$,
\begin{align}\label{Eq:Aval}
    \hat{\Pi}_k=\sum\limits_{l=0}^k \hat{F}_l\left(\eta_k\right)-\sum\limits_{l=0}^{k-1} \hat{F}_l\left(\eta_{k-1}\right)
\end{align}
for $k=1,\dots, K$, and
\begin{align}\label{Eq:AvalLast}
    \hat{\Pi}_{K+1}=1-\sum\limits_{l=0}^K \hat{F}_l\left(\eta_K\right) .    
\end{align}
Here 
\begin{align}\label{Eq:AdjEff}
    \eta_k=\frac{\tau_{\mathrm{m}}-k \tau_{\mathrm{d}}}{\tau_{\mathrm{m}}}
\end{align}
is the adjustment efficiency.

It is worth noting that all equations in this appendix are given for detectors with no losses, i.e., $\eta=1$.
This is a consequence of the fact that we attribute all losses, including the detection losses, to the prepared quantum state.
In order to explicitly include the detection losses in the POVM, one needs to replace $\hat{n}$ with $\eta\hat{n}$ under the sign of the normal ordering.

%%%%%%%%%%%%%%%%%%%%%%%%%%%%%%%%%%%%%%%%%%%%%%%%%%%%%%%%%%%%%%%%%%%%%%%%%%%
%%% Photocounting probabilities with realistic photon-number resolution
%%%%%%%%%%%%%%%%%%%%%%%%%%%%%%%%%%%%%%%%%%%%%%%%%%%%%%%%%%%%%%%%%%%%%%%%%%%

\section{Photocounting probabilities with realistic photon-number resolution}
\label{App:ProbabDistr}

In this appendix, we present the derivation of Eqs.~(\ref{Eq:array}), (\ref{Eq:FunctPulses}), and (\ref{Eq:FAPD}), representing the matrix functional (\ref{mat_f}) for different detection schemes.
First, we consider the functional
\begin{widetext}
\begin{align}
    &F(W,\boldsymbol{m},\boldsymbol{a})=
    \prod\limits_{i=1}^{L} 
    \left( \frac{\partial^2}{\partial \alpha_i \partial \alpha^*_i} \right)^{m_i}
    \exp{\left(a_{i} \frac{ \partial^2}{\partial \alpha_{i} \partial \alpha_{i}} \right)}
    \left. \exp{\left(\frac{1}{2}\boldsymbol{\xi}^{\dagger} (\mathbb{I} - W) \boldsymbol{\xi}\right)}
    \right|_{\boldsymbol{\xi=0}}, 
\end{align}
which is used for further analysis.
Here $\boldsymbol{m}=(m_1,\ldots,m_L)$, $\boldsymbol{a}=(a_1,\ldots,a_L)$, $a_i=[0,1]$, $\boldsymbol{\xi}=(\alpha_1, \ldots, \alpha_L, \alpha_1^{*},\ldots, \alpha_L^{*})^T$, and $W\in \mathcal{C}^{2L \times 2L}$ is a positive-semidefinite matrix, for example, $(\mathbb{I}-A_{\boldsymbol{n}})$. 
Applying the Weierstrass transform,
\begin{align}
    \exp{\left(a_{i} \frac{ \partial^2}{\partial \alpha_{i} \partial \alpha_{i}} \right)}
    \exp\left(\alpha_{i}^*\beta_{i}-\beta_{i}^*\alpha_{i}+|\alpha_{i}|^2
    \right)=
    \frac{1}{1-a_{i}}
    \exp\left(\frac{\alpha_{i}^*\beta_{i}-\beta_{i}^*\alpha_{i}+|\alpha_{i}|^2}
    {1-a_{i}}\right),
\end{align}
we obtain
\begin{align}\label{Eq:useful_A}
    &F(W,\boldsymbol{m},\boldsymbol{a})
    =\frac{1}{\pi^L \sqrt{|W|}}
    \int_{\mathcal{C}^{2L}}d^{2L}\boldsymbol{\beta}
    \\\nonumber
    &\times
    \exp{\left(-\frac{1}{2}\boldsymbol{\beta}^{\dagger} 
    \left[W^{-1} +\diag\left(\frac{a_{i}}{1-a_{i}}\right) \right]
    \boldsymbol{\beta}\right)}
    \prod\limits_{i=1}^{L} 
    \frac{1}{1-a_{i}}
    \left( \frac{\partial^2}{\partial \alpha_i \partial \alpha^*_i} \right)^{m_i}
    \left. \exp\left(\frac{\alpha_{i}^*\beta_{i}-\beta_{i}^*\alpha_{i}+|\alpha_{i}|^2}
    {1-a_{i}}\right)
    \right|_{\boldsymbol{\xi=0}}.
\end{align}
Changing the variables, $\beta_i = \gamma_i \sqrt{1-a_i}$ and $\alpha_i = \mu_i \sqrt{1-a_i}$, and integrating, we arrive at the expression
\begin{align}\label{Eq:useful_f0}
    F(W,\boldsymbol{m},\boldsymbol{a})=
    \left.
    \prod\limits_{i=1}^{L}
    (1-a_i)^{m_i}
    \left( \frac{\partial^2}{\partial \mu_i \partial \mu^*_i} \right)^{m_i}
    \exp{\left(\boldsymbol{\mu}^{\dagger} 
    \left\{\mathbb{I} -  
    \left[\diag\sqrt{1-a_i} W^{-1}\diag\sqrt{1-a_i} +\diag\left(a_{i}\right) \right]^{-1}
    \right\}\boldsymbol{\mu}\right)}
    \right|_{\boldsymbol{\mu=0}},
\end{align}
which can also be given in terms of the Hafnian as
\begin{align}\label{Eq:useful_f}
    F(W,\boldsymbol{m},\boldsymbol{a})=
    \frac{1}{\sqrt{|\mathbb{I} -\diag\sqrt{a_i}(\mathbb{I}-W)\diag\sqrt{a_i}|}}
    \Haf\left(X 
    \left[
    \diag(1-a_i) - \left(W^{-1}
    +\diag\left(\frac{a_{i}}{1-a_i}\right) \right)^{-1}
    \right]_{\boldsymbol{m}}
    \right).
\end{align}
Here the subscript $\boldsymbol{m}$ indicates that the matrix $A_{\boldsymbol{m}}$ is derived from matrix $A$ by retaining solely the rows and columns with indexes $i$ and $i+L$ for which $m_i\neq 0$ and repeating each of them $m_i$ times.

\subsection{Click detectors}
\label{App:ProbabDistr_Click}

Let us consider the derivation of Eq.~(\ref{Eq:array}) for click detectors.
The $Q$ symbol of the POVM $\Pi_n(\alpha)$ can be obtained from the POVM (\ref{Eq:Array}) by replacing $\hat{n}$ by $|\alpha|^2$ under the sign of the normal order.

Substituting it into Eq.~(\ref{mat_f}) yields
\begin{align}\label{Eq:click_gen}
    \mathcal{F}^{\mathrm{click}}_{\boldsymbol{n}}\left[A\right] = 
    \prod\limits_{i=1}^{N} \binom{K}{c_{i}}
    \sum\limits_{k_{1}=0}^{c_{1}} 
    \dots
    \sum\limits_{k_{N}=0}^{c_{N}}
    \prod\limits_{i=1}^{N} \binom{c_{i}}{k_{i}}(-1)^{k_{i}}
    \exp{\left( \frac{c_{i} - k_{i}}{K} \frac{\partial^2}{\partial \alpha_{i} \partial \alpha^*_{i}} \right)}
    \left. \exp{\left(\frac{1}{2}\boldsymbol{\xi}^{\dagger} A_{S(\boldsymbol{n})} \boldsymbol{\xi}\right)}
    \right|_{\boldsymbol{\xi=0}}.
\end{align}
Then we use Eq.~(\ref{Eq:useful_f}) with $\boldsymbol{m}=\boldsymbol{c}$, $a_i=\frac{c_{i} - k_{i}}{K}$, and $W=(\mathbb{I}-A_{S(\boldsymbol{n})})$, which gives
\begin{align}\label{Eq:ken_tru}
    \mathcal{F}^{\mathrm{click}}_{\boldsymbol{n}}\left[A\right]=\Ken[A_{\boldsymbol{n}}] = 
    \prod\limits_{i=1}^{N} \binom{K}{c_{i}}
    \sum\limits_{k_{1}=0}^{c_{1}} 
    \dots
    \sum\limits_{k_{N}=0}^{c_{N}}
    \prod\limits_{i=1}^{N} \binom{c_{i}}{k_{i}}(-1)^{k_{i}}
    \frac{1}{\sqrt{\left|\mathbb{I} - B_{S(\boldsymbol{c})}^{\boldsymbol{k}}\right|}},
\end{align}
where $B_{S(\boldsymbol{c})}^{\boldsymbol{k}}$ is given by Eq.~(\ref{Eq:BSmatrix})

It is also useful to consider the relation between two matrix functionals: $\Ken$ and $\Tor$.
Equation~(\ref{Eq:ken_tru}) can be rewritten as 
\begin{align}\label{Eq:ken_rep}
    \Ken[A_{S(n)}]=\prod\limits_{i=1}^{N} \binom{K}{c_{i}}
    \sum\limits_{k_{1}=0}^{c_{1}-1} 
    \dots
    \sum\limits_{k_{N}=0}^{c_{N}-1}
    \prod\limits_{i=1}^{N} \binom{c_{i}}{k_{i}}(-1)^{k_{i}}
    \sum\limits_{Z \in P([N])}
    \frac{(-1)^{|z|}}{\sqrt{\left|\mathbb{I} - \left(B_{S(\boldsymbol{c})}^{\boldsymbol{k}}\right)_{Z} \right|}}.
\end{align}
Applying here the definition of $\Tor$ (cf.~\cite{Quesada2018}), we can obtain Eq.~(\ref{Eq:KenTor}).
Substituting $K=1$ into Eq. ~(\ref{Eq:ken_rep}), we arrive at Eq.~({\ref{Eq:on/off_Tor}}).

\subsection{Photocounting with SNSPDs}
\label{App:ProbabDistr_SNSPDs}

Consider the scenario of photocounting with SNSPDs, described by the POVM~(\ref{Eq:SNSPD}).
Like in the previous case, the $Q$ symbol of the POVM $\Pi_n(\alpha)$ is obtained by replacing $\hat{n}$ by $|\alpha|^2$ under the sign of the normal order. 
Applying it in Eq.~(\ref{mat_f}), we get
\begin{align}\label{Eq:SPNSDs_gen}
    \mathcal{F}^{\mathrm{SNSPD}}_{\boldsymbol{n}}\left[A\right]=
    \int_{T_{c_{1}}} d^{c_{1}} 
    \boldsymbol{t}_{1}
    \ldots
    &\int_{T_{c_{N}}} d^{c_{N}} 
    \boldsymbol{t}_{N}
    \prod\limits_{j=1}^{N}
    \mathcal{I}_{c_{j}} (\boldsymbol{t}_{j})
    \\\nonumber
    &\left. \times
    \left( \frac{\partial^2}{\partial \alpha_j \partial \alpha^*_j} \right)^{c_j}
    \exp\left([1 - \Xi_{c_{j}}(\boldsymbol{t}_{j})]
    \frac{\partial^2}{\partial \alpha_i \partial \alpha^*_i}\right)
    \exp{\left(\frac{1}{2}\boldsymbol{\xi}^{\dagger} A_{S(\boldsymbol{n})} \boldsymbol{\xi}\right)}
    \right|_{\boldsymbol{\xi=0}}.
\end{align}
Then we use Eq.~(\ref{Eq:useful_f}) with $\boldsymbol{m}=\boldsymbol{c}$, $a_i= [1 - \Xi_{c_{j}}(\boldsymbol{t}_{j})]$, and $W=(1-A_{S(\boldsymbol{n})})$, which leads to Eq.~(\ref{Eq:FunctPulses}).

\subsection{Photocounting with APDs}
\label{App:ProbabDistr_APDS}

Photocounting with the APDs can be considered a particular case of photocounting with the SNSPDs, for which $\tau_\textrm{r}=0$.
We also suppose that the nonmonochromatic light mode has a rectangular envelop [see Eq.~(\ref{Eq:RecEnv})].
Our consideration is restricted by $n_i=\{0,1\}$, i.e., by the collision-free events.
The required POVM elements are given by Eqs.~(\ref{Eq:Aval0}) and (\ref{Eq:Aval}).
The corresponding $Q$ symbols are obtained in the standard way by replacing $\hat{n}$ by $|\alpha|^2$ under the sign of the normal order. 
Applying them in Eq.~(\ref{mat_f}) yields
\begin{align}
    &\mathcal{F}^{\mathrm{APD}}_{\boldsymbol{n}}\left[A\right]=
    \left.\prod\limits_{i=1}^{N}
    \left[
    \left(1 + \eta_1
    \frac{\partial^2}{\partial \alpha_i \partial \alpha^*_i} 
    \right)
    \exp\left([1 - \eta_1]
    \frac{\partial^2}{\partial \alpha_i \partial \alpha^*_i}\right) - 1
    \right]
    \exp{\left(\frac{1}{2}\boldsymbol{\xi}^{\dagger} A_{S(\boldsymbol{n})} \boldsymbol{\xi}\right)}
    \right|_{\boldsymbol{\xi=0}}
    \\\nonumber
    &\left.
    =\sum\limits_{Z \in P[N]}
    (-1)^{|Z|}
    \prod\limits_{i=1}^{|Z|}
    \left(1 + \eta_1
    \frac{\partial^2}{\partial \alpha_i \partial \alpha^*_i} 
    \right)
    \exp\left([1 - \eta_1]
    \frac{\partial^2}{\partial \alpha_i \partial \alpha^*_i}\right)
    \exp{\left(\frac{1}{2}\boldsymbol{\xi}^{\dagger} A_{S(\boldsymbol{n}),Z} \boldsymbol{\xi}\right)}
    \right|_{\boldsymbol{\xi=0}},
\end{align}
where $P[N]$ is explained after Eq.~(\ref{Eq:D_S}).
Utilizing Eq.~(\ref{Eq:useful_f0}), we get
\begin{align}
    &\mathcal{F}^{\mathrm{APD}}_{\boldsymbol{n}}\left[A\right]
    =\sum\limits_{Z \in P[N]}
    \frac{(-1)^{|Z|}}{\pi^N \sqrt{|(\mathbb{I}-A_{S(\boldsymbol{n}),Z})|}}
    \int_{\mathcal{C}^{2N}}d^{2N}\boldsymbol{\beta}
    \exp{\left(-\frac{1}{2}\boldsymbol{\beta}^{\dagger} 
    \left[(\mathbb{I}-A_{S(\boldsymbol{n}),Z})^{-1} +\mathbb{I}\left(\frac{1}{\eta_1}-1\right) \right]
    \boldsymbol{\beta}\right)}
    \\\nonumber
    &\times \prod\limits_{i=1}^{|Z|} 
    \frac{1}{\eta_1}
    \left(1 + \eta_1
    \frac{\partial^2}{\partial \alpha_i \partial \alpha^*_i} 
    \right)
    \left. \exp\left(\frac{\alpha_{i}^*\beta_{i}-\beta_{i}^*\alpha_{i}+|\alpha_{i}|^2}
    {\eta_1}\right)
    \right|_{\boldsymbol{\xi=0}}.
\end{align}
Next, we use the equality
\begin{align}
    \left(1 + \eta_1
    \frac{\partial^2}{\partial \alpha_i \partial \alpha^*_i} 
    \right)
    \left. \exp\left(\frac{\alpha_{i}^*\beta_{i}-\beta_{i}^*\alpha_{i}+|\alpha_{i}|^2}
    {\eta_1}\right)
    \right|_{\alpha_i=0}=
    \left. 2 
    \frac{\partial^2}{\partial \alpha_i \partial \alpha^*_i} 
    \exp\left(\frac{\alpha_{i}^*\beta_{i}-\beta_{i}^*\alpha_{i}}
    {\sqrt{2\eta_1}} +|\alpha_{i}|^2\right)
    \right|_{\alpha_i=0}
\end{align}
and change variables as $\beta = \gamma \sqrt{2\eta_1}$.
This give us the expression
\begin{align}
    &\mathcal{F}^{\mathrm{APD}}_{\boldsymbol{n}}\left[A\right]
    =\sum\limits_{Z \in P[N]}
    \frac{(-1)^{|Z|}}{\pi^N \sqrt{|(\mathbb{I}-A_{S(\boldsymbol{n}),Z})|}}
    \int_{\mathcal{C}^{2N}}d^{2N}\boldsymbol{\gamma}
    \exp{\left(-\frac{1}{2}\boldsymbol{\gamma}^{\dagger} 
    \left[2 \eta_1 (\mathbb{I}-A_{S(\boldsymbol{n}),Z})^{-1} +2\left( 1 - \eta_1 \right) \mathbb{I}\right]
    \boldsymbol{\gamma}\right)}
    \\\nonumber
    &\times \prod\limits_{i=1}^{N} 
    4
    \left.  
    \frac{\partial^2}{\partial \alpha_i \partial \alpha^*_i} 
    \exp\left(\alpha_{i}^*\gamma_{i}-\gamma_{i}^*\alpha_{i}
     +|\alpha_{i}|^2\right)
    \right|_{\boldsymbol{\xi=0}}
\end{align}
Finally, we can see that this expression has a structure similar to Eq.~(\ref{Eq:useful_A}), which is transformed into Eq.~(\ref{Eq:useful_f}).
This leads to Eq.~(\ref{Eq:FAPD}).

%%%%%%%%%%%%%%%%%%%%%%%%%%%%%%%%%%%%%%%%%%%%%%%%%%%%%%%%%%%%%%%%%%%%%%%%%%%
%%% Phase-space simulation
%%%%%%%%%%%%%%%%%%%%%%%%%%%%%%%%%%%%%%%%%%%%%%%%%%%%%%%%%%%%%%%%%%%%%%%%%%%

\section{Phase-space simulation}
\label{App:PhaseSpaceSimm}

In this appendix we briefly sketch the evaluation procedure of the characteristic function $C(k_1,k_2)$ [cf. Eq.~(\ref{Eq:CharF2})] using the methods presented in Refs.~\cite{Opanchuk2018,Drummond2022,Dellios2022, dellios2023a, dellios2023b}.
The corresponding source code is given in the Ancillary Files \cite{supplement}.
First, we generate samples $\Tilde{\boldsymbol{\alpha}}^{\lambda}=\{\Tilde{\alpha}_1^{\lambda},\ldots,\Tilde{\alpha}_M^{\lambda}\}$ and $\Tilde{\boldsymbol{\beta}}^{\lambda}=\{\Tilde{\beta}_1^{\lambda},\ldots,\Tilde{\beta}_M^{\lambda}\}$ for the interferometer inputs, where $\lambda=1,\ldots, E_{\mathrm{S}}$. 
As discussed in Ref.~\cite{Drummond2022}, in the case of the thermalized squeezed states with the thermalization factor $\epsilon_j$ and the squeezing parameter $r_j$, the amplitudes $\Tilde{\alpha}_j^{\lambda}$ and $\Tilde{\beta}_j^{\lambda}$ for the $\lambda$th input mode and the $l$th sampling can be obtained as 
    \begin{align}
        & \Tilde{\alpha}_j^{\lambda}=\delta_{j+} \omega_j^{\lambda}+i \delta_{j-} \omega_{j+M}^{\lambda}, \\
        & \Tilde{\beta}_j^{\lambda}=\delta_{j+} \omega_j^{\lambda}-i \delta_{j-} \omega_{j+M}^{\lambda},
    \end{align}
where $\omega_i^{\lambda}$ are sampled as real Gaussian variables with $\left\langle \omega_i^{\lambda} \omega_j^{\lambda} \right\rangle_P = \delta_{ij}$ and 
    \begin{align}
        \delta_{j \pm}=\sqrt{\frac{\sinh^2(r_j) \pm (1-\epsilon_j) \sinh(r_j)\cosh(r_j)}{2}}.
        \nonumber
    \end{align}
The samples for the output amplitudes are obtained as $\boldsymbol{\alpha}^{\lambda}=U \boldsymbol{\Tilde{\alpha}}^{\lambda}$ and
$\boldsymbol{\beta}^{\lambda}=U^{*} \boldsymbol{\Tilde{\beta}}^{\lambda}$, where $U$ is a unitary matrix, describing the transformation of the coherent amplitudes of the electromagnetic field at the interferometer.
The characteristic function can be estimated as
    \begin{align}
        C(k_1,k_2) \approx 
        \frac{1}{E_{\mathrm{S}}} \sum\limits_{\lambda=1}^{E_{\mathrm{S}}}\Bigg[\prod\limits_{i=1}^M \big[\pi_i(0|\alpha_i^{\lambda},\beta_i^{\lambda})
        +\pi_i(1|\alpha_i^{\lambda},\beta_i^{\lambda}) e^{-i k_1 \theta} + \pi_i(2|\alpha_i^{\lambda},\beta_i^{\lambda}) e^{-i k_2 \theta} \big]\Bigg]
        ,
    \end{align}
where $\pi_i(q|\alpha_i^{\lambda},\beta_i^{\lambda})$ is given by Eq.~(\ref{Eq:POVM_Symb}). 

To optimize the simulation time in the case of $q\leq 2$, we propose the following procedure.
Let us consider the expression
    \begin{align}\label{Eq:PTilde}
        \Tilde{\mathcal{P}} (m)=
        \sum\limits_{p=0}^{p_{\mathrm{max}}}
        \mathcal{P}\left(\mathcal{O}^{m\!\!\!\mod{(M/D)}+2t}_{pJ + t}\right) ,
    \end{align}
where $m=\{ 0, 1, \ldots,  J \lfloor M/D \rfloor - 1  \}$, 
$p_{max} = \lfloor  \{m\!\!\!\mod{(M/D)}\}/(2J) \rfloor$, $t = \lfloor m/(\lfloor M/D \rfloor) \rfloor$, $\lfloor\ldots \rfloor$ is the floor function, and
the numbers $D\in\mathbb{R}_{+}$ and $J\in\mathbb{N}_{+}$ are taken such that the conditions 
    \begin{align}
        &\mathcal{P}\left(\mathcal{O}^{\lfloor M/D\rfloor}_{0}\right) < \frac{1}{E_{\mathrm{S}}},\\
        &\mathcal{P}\left(\mathcal{O}^{\lfloor\left<n\right>\rfloor}_{J}\right) < \frac{1}{E_{\mathrm{S}}}
    \end{align}
are satisfied.
Here $\left<n\right>$ is the mean number of photons at the output.
Expression (\ref{Eq:PTilde}) approximates the orbit probability as
    \begin{align}
        \Tilde{\mathcal{P}} (m)\approx \mathcal{P}(\mathcal{O}^{m_1 + 2m_2}_{m_2}),
    \end{align}
where $m=m_1 + m_2 \lfloor M/D \rfloor$.
To estimate $D$, we use the fact that $\mathcal{P}\left(\mathcal{O}^{\lfloor M/D\rfloor}_{0}\right)$ can be approximated by the probability of getting $\lfloor M/D\rfloor$ photons at the output, $p\left(\lfloor M/D\rfloor\right)$. 
To estimate $J$ in the case of $n^2\approx M$, we use the approximation $\mathcal{P}\left(\mathcal{O}^{\lfloor\left<n\right>\rfloor}_{J}\right) \approx p\left(\lfloor\left<n\right>\rfloor\right)/J!$.

The discrete Fourier transform for $\Tilde{\mathcal{P}} (m)$ can be estimated as
    \begin{align}
        \Tilde{C}(k)& \approx 
        \frac{1}{E_{\mathrm{S}}} \sum\limits_{\lambda=1}^{E_{\mathrm{S}}}\Bigg[\prod\limits_{i=1}^M \big[\pi_i(0|\alpha_i^{\lambda},\beta_i^{\lambda})
        +\pi_i(1|\alpha_i^{\lambda},\beta_i^{\lambda}) e^{-i k \Tilde{\theta}} + \pi_i(2|\alpha_i^{\lambda},\beta_i^{\lambda}) e^{-i k\lfloor M/D\rfloor \Tilde{\theta}} \big]\Bigg]
        .
    \end{align}
    Here $\Tilde{\theta}=2\pi/(J \lfloor M/D \rfloor)$ and $k=\{ 0, 1, \ldots,  J \lfloor M/D \rfloor-1  \}$.
    This function depends only on a single value; thus, its inverse discrete Fourier transform requires significantly fewer computational resources.
    With this approximation, we reduce the run time of the simulation from $O(M^2)$ to $O(JM/D)$.

\end{widetext}

\bibliography{biblio}

\end{document}